\documentclass[12pt,a4paper]
					{article}	
\usepackage[utf8]{inputenc}		
\usepackage[T1]{fontenc}		
\usepackage[english]{babel}		
\usepackage{amsmath}     		
\usepackage{amssymb}     		
\usepackage{amsfonts}   		
\usepackage{amsthm}				
\usepackage[colorlinks=true]
					{hyperref}	
\usepackage{mathrsfs}			
\usepackage{authblk}			
\usepackage{mathscinet}			
\usepackage{dsfont}	
\usepackage[numbers,sort&compress]{natbib}
\usepackage[a4paper,
            left=25mm,      
            right=25mm,     
            top=20mm,       
            bottom=30mm,    
            headheight=15mm,
            headsep=10mm,   
            footskip=15mm]{geometry} 

\hypersetup{	
  pdfstartview = XYZ null null 1.00,
  pdfstartpage = 1
}

\newtheoremstyle{Definition}
	{3pt}			
	{3pt}			
	{}				
	{\parindent}	
	{\scshape}		
	{.}				
	{.5em}			
	{}				
\theoremstyle{Definition}

\title{Pseudo-Riemannian Lie algebras with coisotropic ideals and integrating the Laplace--Beltrami equation on Lie groups}
\author[1]{A.~A.~Magazev\thanks{magazev@omgtu.ru}}
\author[1]{I.~V.~Shirokov\thanks{i\_v\_shirokov@mail.ru}}
\affil[1]{\small Omsk State Technical University, Omsk, Russia}
\date{}

\newcommand{\p}{\partial}					
\newcommand{\g}{\mathfrak{g}}				
\newcommand{\h}{\mathfrak{h}}				
\newcommand{\G}{\mathbf{G}}					
\newcommand{\D}{\mathscr{D}}				
\DeclareMathOperator{\ind}{\mathrm{ind}}	
\newcommand{\defeq}{\mathrel{\mathop:}=}	
\newcommand{\gL}{\mathfrak{g}_L}			
\newcommand{\gR}{\mathfrak{g}_R}			
\newcommand{\C}{\mathbf{C}}					
\newcommand{\pol}{\mathfrak{p}}				
\newcommand{\<}{\langle}					
\renewcommand{\>}{\rangle}					
\newcommand{\Ad}{\mathrm{Ad}}				
\renewcommand{\D}{D}						
\newcommand{\F}{\mathcal{F}}				
\newcommand{\id}{\mathds{1}}				
\newcommand{\Vect}[1]{\mathrm{Vect}(#1)}	
\newcommand{\e}{\mathbf{e}}					
\newcommand{\redDelta}{\widetilde{\Delta}}	
\renewcommand{\H}[2]{\mathrm{H}_{#1}(#2)}	
\newcommand{\R}{\mathbb{R}}					
\DeclareMathOperator{\Ai}{\mathrm{Ai}}		
\DeclareMathOperator{\Bi}{\mathrm{Bi}}		

\begin{document}
\maketitle
\selectlanguage{english}

\begin{abstract}
We identify a class of left-invariant pseudo-Riemannian metrics on Lie groups for which the Laplace--Beltrami equation reduces to a first-order PDE and admits exact solutions. The defining condition is the existence of a commutative ideal $\mathfrak{h}$ in the Lie algebra $\mathfrak{g}$ whose orthogonal complement satisfies $\mathfrak{h}^\perp\subseteq\mathfrak{h}$. Using the noncommutative integration method based on the orbit method and generalized Fourier transforms, we reduce the Laplace--Beltrami equation to a first-order linear PDE, which can then be integrated explicitly. The symmetry of the reduced equation gives rise, via the inverse transform, to nonlocal symmetry operators for the original equation. These operators are generically integro-differential, contrasting with the polynomial symmetries appearing in previously studied classes. The method is illustrated by two examples: the Heisenberg group $\mathrm{H}_3(\mathbb{R})$ with a Lorentzian metric and a four-dimensional non-unimodular group with a metric of signature $(2,2)$. In the latter, classical separation of variables is not directly applicable, yet the noncommutative approach yields explicit solutions and reveals the predicted nonlocal symmetry.
\end{abstract}

\tableofcontents

\section*{Introduction}
\addcontentsline{toc}{section}{Introduction}

Systems, both classical and quantum, that possess hidden symmetries are of significant interest due to their enhanced integrability and non-trivial dynamics.
Lie groups endowed with left-invariant pseudo-Riemannian metrics play a distinguished role in this context, as they provide a natural algebraic framework for analyzing such systems. 
This framework allows deep problems of geometry and dynamics to be translated into more tractable problems within Lie algebra.

In differential geometry, the study of left-inva\-ri\-ant metrics naturally branches into several profound and interconnected areas. 
A primary focus is the study of the pseudo-Riemannian structures themselves, including key properties like sectional curvature and the Einstein equations. 
It is important to note, however, that the majority of existing results in this area concern Riemannian and Lorentzian signatures~\cite{Mil76, CheLia12, AndTor20, Nom79}, while metrics of general indefinite signature remain far less explored.
Concurrently, the central problems of dynamics and analysis on these manifolds are of significant independent interest. 
This includes the problem of integrating geodesic and magnetic geodesic flows, which describe the free particle motion~\cite{AbrMar78, MisFom78, MagShiYur08, InoMun23}, as well as the study of the associated Laplace--Beltrami operator.
Its spectral properties, determined by the left-invariant metric, yield profound information about the geometry of the group~\cite{BeeMil77, Ura79, RicMul88, Lau23}. 
Moreover, exact solutions to the Laplace--Beltrami equation, including on Lie groups, are of direct physical relevance, as this equation governs the dynamics of scalar fields in various classical and quantum field theories~\cite{Mag12, BreShiMag11, BarMikShi02, Obu23, UglShi07, Bre10}.
Such solutions play a crucial role, for example, in quantum field theory on curved backgrounds and in the analysis of cosmological models~\cite{Obu24, Obu24b, Obu25}.

The present work aims to identify and describe a special class of left-invariant pseudo-Riemannian metrics on Lie groups that admit exact solutions to the associated Laplace--Beltrami equations. 
Specifically, our primary objective is to establish structural conditions on the Lie algebra and the metric under which this second-order PDE reduces completely to a \textit{first-order} equation, thereby becoming explicitly integrable.

While the Laplace--Beltrami equation, $\Delta \psi = E \psi$, is fundamental in both Riemannian geometry (as an eigenvalue problem), and Lorentzian geometry (as the Klein--Gordon equation), its explicit solvability on a general Lie group remains a challenging problem. 
The natural first-order symmetries---the right-invariant vector fields---commute with a left-invariant Laplacian $\Delta$, but typically do not form a complete set for integration. 
Therefore, a complete solution generally requires higher-order symmetries or additional algebraic structure. 
Examples of such metrics with additional symmetries are bi-invariant metrics or, for instance, left-invariant metrics constructed by the argument shift method~\cite{MisFom78}.
The generalization of this method to the quantum context is discussed in~\cite{IkeSha24}. 
Beyond these, there are several other powerful approaches to constructing left-invariant metrics with hidden symmetries. Key examples include the method of subalgebra chains~\cite{Thi81}, the Lax pair formalism~\cite{BraAraBlo25}, and techniques utilizing Lie algebra filtrations~\cite{Bog84, JovSukVuk23}. 
A common feature of these methods is that the additional integrals of motion they yield are typically polynomial in the canonical momenta. 
In contrast to these approaches, the class of left-invariant metrics studied in the present work gives rise to additional integrals of motion that are not necessarily polynomial in the momenta. 
Consequently, for the Laplace--Beltrami equation, the corresponding symmetry operators are nonlocal rather than differential operators of finite order. We note that in the Lorentzian signature, the condition we introduce reduces to the well-known class of almost abelian Lie algebras~\cite{Gor2020, Ave2022, FinPar2023}, while for higher indefinite signatures we obtain a substantially broader family, including non-unimodular examples.

This paper is organized as follows.
In Section~\ref{sec:1}, we recall the basic facts about left-invariant pseudo-Riemannian metrics on Lie groups and give a formula for the Laplace--Beltrami operator expressed in terms of left-invariant vector fields. We note that the basis in the corresponding Lie algebra is not assumed to be orthonormal.
In Section~\ref{sec:2}, we describe the general framework underlying our approach to the integration of linear differential equations on Lie groups. This is the so-called \textit{method of non-commutative integration} of linear differential equations, suggested by A. Shapovalov and I. Shirokov~\cite{ShaShi95}. 
The method exploits the symmetry of the Laplace–Beltrami equation generated by left translations and, using orbit method and harmonic analysis on the Lie group, reduces the original equation to one with fewer independent variables.
In the general case, this reduced equation remains a second-order partial differential equation. 
In section~\ref{sec:3}, we identify a specific class of metrics for which this reduced equation simplifies further, becoming a first-order equation and thereby allowing for explicit integration.
These metrics are related to the presence of coisotropic commutative ideals within pseudo-Riemannian Lie algebras and, in non-trivial cases, are non-Riemannian.
A key feature of the Laplace–Beltrami equation for these metrics is the emergence of nonlocal symmetry operators, which are realized as integro-differential operators.
Section~\ref{ExaSec} contains explicit constructions that substantiate the general framework. We first analyze the three-dimensional Heisenberg group~$\H{3}{\R}$ endowed with a Lorentzian metric satisfying the coisotropic ideal condition. This example illustrates the equivalence between the noncommutative approach and classical separation of variables when the latter is applicable. 
Subsequently, we consider a four-dimensional non-unimodular Lie group (Mubarakzyanov class $\g_{4,7}$) equipped with a metric of signature $(2,2)$. 
In this case, the absence of a three-dimensional commutative subalgebra means that separation of variables, if possible at all, would require the construction of higher-order symmetry operators---a significantly more challenging problem. 
We construct the general solution via the noncommutative Fourier transform and identify the corresponding nonlocal symmetry operators, thereby confirming the distinctive properties predicted in Section~\ref{sec:3}.

\section{Left-invariant Laplace--Beltrami equations}
\label{sec:1}
This section offers a concise overview of left-invariant pseudo-Riemannian metrics on Lie groups and their associated Laplace--Beltrami operators. 
For a comprehensive treatment of this material, the reader is referred to the standard references~\cite{War83, Hel78}.

Let $G$ be an $n$-dimensional connected real Lie group. 
The group $G$ acts smoothly on itself via left and right translations, defined respectively by the maps $L_y \colon G \to G$, $L_y(x) = y x$, and $R_y \colon G \to G$, $R_y(x) = xy$, for each $y \in G$.
A vector field $\zeta \in \Vect{G}$ is called \textit{left-invariant} if it satisfies $dL_y(\zeta) = \zeta$ for all $y \in G$. 
Similarly, $\zeta$ is \textit{right-invariant} if $dR_y(\zeta) = \zeta$ for all $y \in G$.
The space of all left-invariant (respectively right-invariant) vector fields on $G$ is closed under the Lie bracket $[\cdot, \cdot]$ and thus forms a Lie algebra. 
This Lie algebra, referred to as the \textit{left} (respectively \textit{right}) \textit{Lie algebra} of $G$, is denoted by $\gL$ (respectively $\gR$). 
Since right and left multiplication maps commute, i.~e., $L_x \circ R_y = R_y \circ L_x$ for all $x, y \in G$, the corresponding left-invariant vector fields are Lie-bracket-commuting with the right-invariant vector fields.
Thus, for the Lie algebras $\gL$ and $\gR$ we obtain
\begin{equation*}
[\gL, \gL] \subseteq \gL,\quad
[\gR, \gR] \subseteq \gR,\quad
[\gL, \gR] = 0.
\end{equation*}  

There is a canonical vector space isomorphism $\xi \colon T_e G \to \gL$ between the tangent space $T_e G$ of the Lie group $G$ at the identity $e \in G$ and its left Lie algebra $\gL$. 
This isomorphism is given by $\xi_X(x) = dL_x(X)$, where $\xi_X$ is the image of a tangent vector $X \in T_e G$ under $\xi$, $x \in G$.
The Lie algebra~$\g$ of the Lie group $G$  is identified with the tangent space $T_e G$, and the isomorphism $\xi$ allows us to transfer the Lie bracket from $\gL$ to $\g$ by requiring
\begin{equation}
\label{xi_comm}
[\xi_X, \xi_Y] = \xi_{[X,Y]_\g},\quad
X, Y \in \g.
\end{equation} 
Thus, by definition, $\g$ and $\gL$ are isomorphic as Lie algebras.

Analogously one can consider the isomorphism $\eta \colon T_e G \to \gR$ between the tangent space $T_e G$ and the right Lie algebra $\gR$. This isomorphism is given by the formula $\eta_X(x) = dR_x(X)$, where $X \in T_e G \simeq \g$, $x \in G$. 
However, the vector space isomorphism $\eta$ is an anti-isomorphism of Lie algebras:
\begin{equation}
\label{eta_comm}
[\eta_X, \eta_Y] = - \eta_{[X,Y]_\g},\quad
X, Y \in \g.
\end{equation}

For further discussion, we fix a basis $\e = \{ e_1, \dots, e_n \}$ for the Lie algebra $\g$. 
The Lie bracket $[\cdot,\cdot]_\g$ is completely determined by its action on basis elements, which can be expressed in terms of the \textit{structure constants} $C_{ij}^k \in \mathbb{R}$ via the commutation relations:
\begin{equation*}
[e_i, e_j]_\g = \sum_{k = 1}^n C_{ij}^k e_k,\quad
i,j = 1, \dots, n.
\end{equation*}
We denote by $\xi_i$ and $\eta_i$ the left-invariant and right-invariant vector fields on $G$ extending $e_i$, respectively, i.e.,
\begin{equation*}
\xi_i(x) \defeq dL_x(e_i),\quad
\eta_i(x) \defeq dR_x(e_i),\quad
x \in G,\quad 
i = 1, \dots, n.
\end{equation*}
It follows from Eqs.~\eqref{xi_comm} and \eqref{eta_comm} that
\begin{equation}
\label{xi_eta_comms}
[\xi_i, \xi_j] = \sum_{k = 1}^n C_{ij}^k \xi_k,\quad
[\eta_i, \eta_j] = - \sum_{k = 1}^n C_{ij}^k \eta_k,\quad
[\xi_i, \eta_j] = 0,\quad
i,j = 1, \dots, n.
\end{equation}
 
Now we consider left-invariant pseudo-Riemannian metrics on Lie groups.
A pseudo-Riemannian metric $g$ on $G$ is said to be \textit{left-invariant} if it is preserved under left translations:
\begin{equation*}
(L_z)^* g = g\quad\text{for all}\quad z \in G.
\end{equation*}
A Lie group $G$ endowed with such a metric $g$ is referred to as a \textit{pseudo-Riemannian Lie group} and is denoted by the pair $(G,g)$.
There exists a canonical bijective correspondence between left-invariant pseudo-Riemannian metrics $g$
on $G$ and non-degenerate symmetric bilinear forms on the Lie algebra~$\g$. 
Explicitly, this correspondence is given by
\begin{equation}
\label{g_zeta}
g(\zeta_1, \zeta_2) = \G( dL_{x^{-1}}(\zeta_1), dL_{x^{-1}}(\zeta_2)),\quad
\zeta_1, \zeta_2 \in T_x G.
\end{equation}
A Lie algebra $\g$ equipped with a non-degenerate symmetric bilinear $\G$ form is termed a \textit{pseudo-Riemannian Lie algebra} and is denoted by $(\g, \G)$.

If $\e = \{ e_1, \dots, e_n \}$ is a basis  of the Lie algebra $\g$, then the left-invariant metric $g$ is determined by its values on this basis, which we denote by the constant matrix $\G_{ij} \defeq g(\xi_i, \xi_j)|_e = \G(e_i, e_j)$. 
Here, $\xi_i$ are the left-invariant vector fields corresponding to $e_i$.
Consequently, the metric can be expressed globally as
\begin{equation}
\label{LI_metric}
g = \sum_{i,j = 1}^n \G_{ij}\, \omega^i \otimes \omega^j,
\end{equation}
where the 1-forms $\omega^i$ constitute the dual coframe to the vector fields $\xi_i$, satisfying $\omega^i(\xi_j) = \delta^i_j$.
A key property of any left-invariant metric is that the right-invariant vector fields $\eta_i$ are Killing vector fields. This follows from the fact that left-invariant and right-invariant vector fields commute, $[\xi_i, \eta_j] = 0$. Therefore, the Lie derivative of the metric along any right-invariant field vanishes:
\begin{equation*}
\mathscr{L}_{\eta_i} g = 0, \quad \text{for } i = 1, \dots, n.
\end{equation*}
This means that right-translations are isometries of $g$.

It is important to note, however, that the right-invariant fields $\eta_i$ do not necessarily generate the entire algebra of Killing vector fields. 
The full isometry group $\mathrm{Iso}(G,g)$ of the pseudo-Riemannian manifold $(G, g)$ can be—and often is—larger than the group $G$ itself. 
For examples, see \cite{OchTak76, Shi97, CosReg22}.

Let $\Delta$ denote the Laplace--Beltrami operator (or Laplacian) associated with the left-invariant metric~\eqref{LI_metric}. A key advantage of working on a Lie group with a left-invariant metric is that $\Delta$ admits a global, coordinate-free expression.
In terms of the left-invariant vector fields $\xi_i$ that form a basis for the Lie algebra $\g$, the Laplacian is given by:
\begin{equation}
\label{xiLaplacian}
\Delta = \sum_{i,j=1}^n \G^{ij} \, \xi_i \xi_j + \sum_{i=1}^n \C^i \xi_i,
\end{equation}
where $\G^{ij}$ are the components of the inverse of the matrix $(\G_{ij})$, and the coefficients $\C^i$ are given by
\begin{equation*}
\C^i \defeq -\sum_{j,k=1}^n \G^{ij} C_{jk}^k.
\end{equation*}
The vector field term $\sum \C^i \xi_i$ arises from the divergence of the metric's volume form with respect to the left-invariant frame. 
In fact, for a unimodular Lie group $G$, this term vanishes, and the Laplacian simplifies to the double sum:
\begin{equation*}
\Delta = \sum_{i,j=1}^n \G^{ij} \, \xi_i \xi_j \quad \text{(for unimodular $G$)}.
\end{equation*}
For comparison, the local coordinate expression for the Laplace--Beltrami operator in a chart $x = (x^1, \dots, x^n)$ is:
\begin{equation*}
\Delta = \frac{1}{\sqrt{|\bar{g}(x)|}} \sum_{i,j=1}^n \frac{\p}{\p x^i} \left( g^{ij}(x) \sqrt{|\bar{g}(x)|} \, \frac{\p}{\p x^j} \right),
\end{equation*}
where $g_{ij}(x) = \sum_{k,l} \G_{kl}\, \omega^k_i(x) \omega^l_j(x)$ is the coordinate representation of the metric and $\bar{g}(x) = \det \| g_{ij}(x) \|$. While this form is general, expression \eqref{xiLaplacian} is more natural in this context because it is formulated entirely in terms of the globally defined objects $\xi_i$.

The central object of our study is the left-invariant \textit{Laplace--Beltrami equation}:
\begin{equation}
\label{LBeq}
\Delta \psi = E \psi,
\end{equation}
where $\Delta$ is the Laplacian of a left-invariant metric, $E$ is a real parameter, and $\psi$ is a function on the group $G$.
The physical and mathematical interpretation of this equation depends on the signature of the metric:
\begin{itemize}
\item In the Riemannian case, Eq.~\eqref{LBeq} is a standard eigenvalue problem for the Laplacian, where $E$ and $\psi$ represent eigenvalues and eigenfunctions, respectively~\cite{BeeMil77, Lau20, Lau23, KenKeiNob93}.
\item In the Lorentzian case, the same equation is known as the Klein--Gordon equation, a fundamental wave equation in relativistic physics. Here, the parameter $E$ is interpreted (up to a sign) as the square of the mass, $m^2$~\cite{BarMikShi02, BreShiMag11}.
\end{itemize}
The study of left- or right-invariant Klein--Gordon equations on Lie groups has been undertaken in~\cite{BarMikShi02, BreShiMag11}. 
Furthermore, scalar wave equations on Lorentzian manifolds that admit a simply transitive group of motions can be naturally reformulated as left-invariant Klein--Gordon equations of the form~\eqref{LBeq}~\cite{Mag12, Obu23}.

Our paper focuses on the symmetries of the left-invariant Laplace--Beltrami equation and their application to finding exact solutions.
An ``obvious'' set of symmetry operators for any equation of the form \eqref{LBeq} is given by the right-invariant vector fields $\eta_i$.
The linear span of these vector fields forms the right Lie algebra $\gR$, which is anti-isomorphic to the Lie algebra~$\g$. 
Crucially, when treated as first-order differential operators, the $\eta_i$ commute with the left-invariant Laplacian~$\Delta$:
\begin{equation*}
[\Delta, \eta_i] = 0,\quad
i = 1, \dots, n,
\end{equation*}
as a direct consequence of Eq. \eqref{xi_eta_comms}.
While the algebra $\gR$ provides a starting point, it is generally insufficient for constructing the general solution to~\eqref{LBeq}. 
A complete solution requires finding additional symmetry operators beyond those in $\gR$. 
The existence of these additional symmetries depends on the specific Lie algebra $\g$ and the chosen left-invariant metric.

As a key step towards solving Eq.~\eqref{LBeq}, we employ $\gR$ to perform the \textit{noncommutative reduction} of \eqref{LBeq}. 
This procedure leverages the symmetry to reduce the original PDE to one with fewer independent variables. 
The following section demonstrates how this reduction can be effectively implemented using the framework of harmonic analysis on Lie groups.

Before proceeding to the description of our method for constructing exact solutions to Eq.~\eqref{LBeq}, we make an important remark regarding the functional setting.
In this paper, we focus on the algebraic aspects of constructing solutions. 
The solutions we obtain are to be understood in the sense of distributions (or generalized functions) on $G$, unless specified otherwise. 
The question of their belonging to specific function spaces (e.~g., $L^2(G)$, Schwartz spaces, etc.) depends on the choice of the arbitrary functions appearing in the construction and on the asymptotic conditions imposed; this functional-analytic aspect is not addressed here and is left for future investigation.

\section{Harmonic analysis on Lie groups and the noncommutative reduction}
\label{sec:2}

Harmonic analysis on Lie groups generalizes classical Fourier analysis from Euclidean space to group manifolds. 
This framework is particularly powerful for solving differential equations with group symmetries, such as those arising in quantum physics.
The standard technique for this involves constructing simultaneous eigenfunctions of a complete set of commuting symmetry operators~(see, e.g.,~\cite{BarRac86}).
However, this often requires high-order symmetries, which can be difficult to construct and whose corresponding eigenfunctions are challenging to compute.
An alternative is provided by the orbit method~\cite{Kir76, Kir94, Kir04},  which serves as a foundation for a \textit{noncommutative integration} of linear PDEs on Lie groups and homogeneous spaces. 
This framework, developed by I.~Shirokov and co-authors~\cite{BreShiMag11, BarMikShi01, GonShi09, BagBalGitShi02, UglShi07, Shi04}, offers a distinct advantage: it does not rely on involving high-order symmetry operators. Instead, the method works directly with non-Abelian algebras of first-order symmetries, as will be demonstrated in the following sections.

Before discussing the technical details of harmonic analysis on Lie groups, we first provide a brief overview of key results concerning Lie algebra polarizations. 
For a comprehensive treatment see~\cite{Dix96, TauYu05, Kir04}.

Let $\g^*$ denote the dual space of the Lie algebra $\g$. The \textit{Kirillov form} associated to an element $\lambda \in \g^*$ is the skew-symmetric bilinear form $B_\lambda \colon \g \times \g \to \mathbb{R}$ defined by
\begin{equation*}
B_\lambda(X,Y) = \< \lambda, [X,Y]_\g \>,\quad
X,Y \in \g,
\end{equation*}
where $\<\cdot, \cdot\>$ is the canonical pairing between~$\g^*$ and $\g$.
The kernel of the Kirillov form, denoted $\g_\lambda$, is a subalgebra of $\g$ known as the \textit{annihilator} (or \textit{stabilizer subalgebra}) of $\lambda$:
\begin{equation*}
\g_\lambda = \{ X \in \g \colon B_\lambda(X,Y) = 0\ \text{for all}\ Y \in \g \}.
\end{equation*}
An element $\lambda \in \g^*$ is called \textit{regular} if the dimension of $\g_\lambda$ is minimal, and \textit{singular} otherwise.

The \textit{index} of a Lie algebra $\g$ is defined as the minimal dimension of an annihilator:
\begin{equation*}
\ind \g = \min_{\lambda \in \g^*} \dim \g_\lambda.
\end{equation*} 
This invariant plays a central role in our subsequent analysis.

For a reductive Lie algebra, the index coincides with the rank, which is the dimension of a Cartan subalgebra. 
At the other extreme, a Lie algebra is called \textit{Frobenius} if $\ind \g = 0$. 
This condition is equivalent to the existence of an element $\lambda \in \g^*$ for which the Kirillov form~$B_\lambda$ is non-degenerate.

A subalgebra $\pol \subset \g$ is called a \textit{polarization} at $\lambda \in \g^*$ if it is a maximal isotropic (i.~e., subordinate to $\lambda$) subalgebra. 
This implies two conditions:
\begin{enumerate}
\item It is subordinate to $\lambda$: 
\begin{equation}
\label{SubCond}
\< \lambda, [\pol,\pol]_\g \> = 0.
\end{equation}
\item It has maximal possible dimension, which for a subordinate subalgebra is $\dim \pol = ( \dim \g + \dim \g_\lambda )/2$.
\end{enumerate}
We denote the set of polarizations of $\g$ at $\lambda$ by $\mathfrak{P}(\lambda)$. 
In the important case where $\lambda$ is regular, the dimension of its annihilator $\g_\lambda$ is minimal and equal to the index $\ind \g$. 
Consequently, the dimension of any polarization $\pol \in \mathfrak{P}(\lambda)$ is given by:
\begin{equation}
\label{dimp}
\dim \pol = \frac{1}{2} \left ( \dim \g + \ind \g \right ).
\end{equation}

The existence theory for Lie algebra polarizations is now well-developed, largely due to its central role in the orbit method for geometric quantization~\cite{Kir01, Woo92}.
A comprehensive overview of these results can be found in Dixmier's monograph~\cite{Dix96}.
A key finding, for instance, is that a polarization exists for any regular element $\lambda \in \g^*$ when $\g$ is a complex Lie algebra.
For real Lie algebras, however, polarizations may not exist even for regular functionals. A standard technique to overcome this obstacle is to complexify the Lie algebra, working with $\g^\mathbb{C} = \g \otimes \mathbb{C}$, and then select a polarization within $\g^\mathbb{C}$ (where functionals in $\g^*$ are extended by linearity). This approach necessitates certain technical modifications related to the complexification procedure.

Since the class of Lie algebras considered in this paper always admits real polarizations (specifically, those with a coisotropic commutative ideal, as introduced in Section~\ref{sec:3}), we will not deal with the complex case. 
The interested reader may consult~\cite{Shi00, Mag12, GonShi09, BarMikShi01} for a comprehensive treatment of such scenarios.

We now introduce an important class of Lie algebra representations: the so-called \textit{$\lambda$-representations}.

Let $\lambda \in \g^*$ be a regular element and let $\pol$ be a polarization at~$\lambda$.
The Eq.~\eqref{SubCond} implies that the functional $\lambda$ defines a one-dimensional representation $t^\lambda$ of the subalgebra $\pol$ by
\begin{equation*}
t^\lambda(X) = i \< \lambda, X \>,\quad
X \in \pol.
\end{equation*}
Consider the space $L^\lambda(G,\pol) \subset C^\infty(G)$ consisting of smooth functions $\psi \colon G \to \mathbb{C}$ that satisfy the system of linear differential equations:
\begin{equation}
\label{sys}
\eta_X \psi = t^\lambda(X) \psi,\quad
X \in \pol,
\end{equation}
where $\eta_X \psi$ denotes the action of the right-invariant vector field corresponding to $X$ on $\psi$.
The system~\eqref{sys} is consistent (i.e., solvable) because the operators $\eta_X$ ($X \in \pol$) form a commutative family when restricted to $L^\lambda(G,\pol)$.
This follows from the identity $\eta_{[X,Y]\mathfrak{g}} = [\eta_X, \eta_Y]$ and condition~\eqref{SubCond}, which states that $\pol$ is subordinate to $\lambda$, thereby ensuring all integrability conditions are met.

One of the important results from Kirillov's monograph~\cite{Kir76} states that if there exists a closed subgroup $P \subset G$ with Lie algebra $\pol$, then the space $L^\lambda(G,\pol)$ is isomorphic to $C^\infty(Q)$, where $Q = G/P$ is the corresponding homogeneous space. 
The dimension of this space is $\dim Q = \dim \g - \dim \pol$.
Using the expression for $\dim \mathfrak{p}$ from Eq.~\eqref{dimp}, we find:
\begin{equation}
\label{dimQ}
\dim Q = \frac{1}{2} \left ( \dim \g - \ind \g \right ). 
\end{equation}

The subspace $L^\lambda(G,\pol) \subset C^\infty(G)$ is invariant under the action of left-invariant vector fields $\xi_X$, $X \in \g$.
This follows because left- and right-invariant vector fields commute, ensuring that the action of $\xi_X$ preserves the system of equations~\eqref{sys}.
Using the isomorphism $\phi \colon L^\lambda(G,\pol) \to C^\infty(Q)$, we can transfer this action to the space of functions on the homogeneous space. 
This defines a representation $\ell$ of the Lie algebra on $C^\infty(Q)$ by
\begin{equation*}
\ell_X = \phi \circ \xi_X \circ \phi^{-1},\quad
X \in \g.
\end{equation*}
This representation is called a \textit{$\lambda$-representation} of the Lie algebra~$\g$.
A fundamental property of these representations is that they are operator-irreducible: any bounded operator on $C^\infty(Q)$ that commutes with all operators $\ell_X$ is a scalar multiple of the identity operator.

By construction, the regular elements $\lambda \in \g^*$ label the $\lambda$-representations of the Lie algebra~$\g$. 
However, this labeling is not unique: different labels can yield equivalent representations. 
The equivalence is governed by the coadjoint action of the group~$G$.

Let $\Ad \colon G \to \mathrm{Aut}(\g)$ be the adjoint representation of the group $G$,  defined by $\Ad_z = dL_z \circ dR_{z^{-1}}$ for $z \in G$.
Its dual, the coadjoint representation $\Ad^* \colon G \to \mathrm{Aut}(\g^*)$, acts on~$\g^*$.
Now, suppose that $\lambda_2 = \Ad^*(z) \lambda_1$ for some $z \in G$ and regular elements $\lambda_1, \lambda_2 \in \g^*$. 
If $\pol_1 \in \mathfrak{P}(\lambda)$ is a polarization at~$\lambda_1$, then $\pol_2 = \Ad(z)( \pol_1 )$ is a polarization at~$\lambda_2$.
Consequently, the function spaces $L^{\lambda_1}(G, \pol_1)$ and $L^{\lambda_2}(G, \pol_2)$ are related by the left translation operator $T^L_z$, where $(T^L_z \psi)(x) = \psi(z^{-1}x)$:
\begin{equation*}
L^{\lambda_2}(G, \pol_2) = T^L_z L^{\lambda_1}(G, \pol_1).
\end{equation*}
Let $P_1$ and $P_2$ be the connected subgroups of $G$ with Lie algebras $\pol_1$ and $\pol_2$, respectively.
Since $P_2 = z P_1 z^{-1}$, the homogeneous spaces $Q_1 = G/P_1$ and $Q_2 = G/P_2$ are isomorphic. 
Denote the corresponding isomorphisms by $\phi_1 \colon L^{\lambda_1}(G,\pol_1) \to C^\infty(Q_1)$ and $\phi_2 \colon L^{\lambda_2}(G, \pol_2) \to C^\infty(Q_2)$.
We now define the intertwining map $\rho \colon C^\infty(Q_1) \to C^\infty(Q_2)$ by
$$
\rho = \phi_2 \circ T^L_z \circ \phi^{-1}_1.
$$
A direct computation shows that $\rho$ is a bijective $\g$-homomorphism. That is, for all $X \in \g$,
\begin{equation*}
\ell^{(2)}_{X} \circ \rho = 
\rho \circ \ell^{(1)}_X,\quad
X \in \g,
\end{equation*}
where $\ell^{(1)}_X = \phi_1 \circ \xi_X \circ \phi_1^{-1}$ and $\ell^{(2)}_X = \phi_2 \circ \xi_X \circ \phi_2^{-1}$ are the respective $\lambda$-representations.
Thus, the $\lambda$-representations associated with coadjoint-equivalent elements $\lambda_1, \lambda_2 \in \g^*$ are equivalent.

We now introduce a parameterization of the regular coadjoint orbits in $\g^*$ to isolate inequivalent $\lambda$-representations.
This is achieved via a map $\sigma \colon \mathcal{J} \to \g^*$, which provides a one-to-one correspondence between an open subset $\mathcal{J} \subset \mathbb{R}^{\ind \g}$ and representatives of the orbits. It is important to note that $\mathcal{J}$ may not be a connected subset of $\mathbb{R}^{\ind \g}$.

Let $\e = \{ e_1, \dots, e_n \}$ be a basis of the Lie algebra $\g$, and let $\ell_i \defeq \ell_{e_i}$ denote the $\lambda$-representation operator corresponding to $e_i$.
On the homogeneous manifold $Q = G / P$, equipped with local coordinates $(q^1, \dots, q^m)$, where $m = \dim Q$, the operators $\ell_i$ are realized as inhomogeneous first-order differential operators acting on $C^\infty(Q)$:
\begin{equation*}
\ell_i(q, \p_q; J) = \sum_{A = 1}^m \zeta_i^A(q) \frac{\p}{\p q^A} + \chi_i(q; J),\quad
i = 1, \dots, n.
\end{equation*}
These operators satisfy the commutation relations of $\g$:
\begin{equation*}
[\ell_i, \ell_j] = \sum_{k = 1}^n C_{ij}^k \ell_k.
\end{equation*}
The vector fields $\zeta_i = \sum_{A=1}^{m} \zeta_i^A(q) \frac{\partial}{\partial q^A}$ are the generators of the $G$-action on $Q = G/H$.
The functions $\chi_i(q; J)$, which depend on the $\ind \g$-valued parameter $J = (J_1, \dots, J_{\ind \g}) \in \mathcal{J}$, are such that the linear map $e_i \mapsto \chi_i$ defines a $2$-cocycle on the Lie algebra $\g$ with values in the $\g$-module $C^\infty(Q)$~\cite{Shi00, Mil95}.
Finally, since the $\lambda$-representations are operator-irreducible, every Casimir operator acts as a scalar multiple of the identity. 
This scalar depends on the parameter $J$:
\begin{equation*}
K(\ell(q, \p_q; J)) = \varkappa(J) \cdot \id\quad\text{for all } K \in Z(U(\g)).
\end{equation*}

To equip the space of functions on $Q$ with a Hilbert space structure, we introduce a measure $d\mu(q)$ on the homogeneous space $Q$. 
This measure is not fixed a priori but is chosen so that the operators $\ell_i$ become skew-Hermitian with respect to the corresponding inner product
\begin{equation}
\label{InnProd}
(\varphi, \psi)_Q = \int_Q \overline{\varphi(q)} \psi(q) d\mu(q). 
\end{equation}
For the classes of Lie groups and polarizations considered in this paper, such a measure can always be constructed explicitly (see Section~\ref{ExaSec} for concrete examples).
Thus, with respect to this inner product, we make the following assumptions regarding the operators~$\ell_i$:
\begin{itemize}
\item They are skew-Hermitian on a common, dense, and invariant domain $\mathcal{D} \subset L^2(Q, d\mu)$.
\item  Every vector in the domain $\mathcal{D}$ is an \textit{analytic} vector for each operator $\ell_i$.
\item The operators $\ell_i$ form a representation of the Lie algebra $\g$ on $\mathcal{D}$.
\end{itemize}

The significance of $\lambda$-representations is their role in formulating harmonic analysis on Lie groups.
A crucial step in this program is to lift the $\lambda$-representation of the Lie algebra $\g$ to a representation of the corresponding connected Lie group~$G$.

Let $T^J$ denote the local representation of the Lie group $G$, obtained by lifting the $\lambda$-representation of the Lie algebra $\g$ that is associated with the dual element $\sigma(J) \in \g^*$.
This lift is defined in a neighborhood $U_e$ of the identity by the requirement that its infinitesimal generator agrees with the $\lambda$-representation operator:
\begin{equation}
\label{loc_rep}
\frac{d}{dt}\, T^J(e^{tX}) \varphi \, \big|_{t = 0} = \ell_X \varphi,\quad\text{for all }
X \in \g;\quad
\varphi \in \mathcal{D} \subset L^2(Q, d\mu(q)).
\end{equation}
By definition, the action of the representation $T^J(x)$ for 
$x \in U_e$ is given by an integral operator:
\begin{equation}
\label{defD}
T^J(x) \varphi (q) = \int_Q \D^J_{qq'}(x) \varphi(q') d\mu(q'),\quad
x \in U_e,
\end{equation}
where the kernel functions $\D^J_{qq'}(x)$ are the matrix elements of $T^J(x)$ with respect to the generalized basis of delta functions $\delta(q,\cdot)$.
Here, the delta distribution $\delta(q,q')$ on $Q$ is defined by the property
\begin{equation*}
\int_Q \delta(q,q') \varphi(q') d\mu(q') = \varphi(q)\quad
\forall\ \varphi \in \mathcal{D} \subset L^2(Q, d\mu(q)).
\end{equation*}
In general, the matrix elements $\D^J_{qq'}(x)$ are distributions on~$G$.
Since $T^J(x)$ is a local group representation, the kernel $\D^J$ satisfies the following relations for $x,y \in U_e$ such that $xy \in U_e$:
\begin{equation}
\label{D(xy)}
\D^J_{qq'}(xy) = \int_Q \D^J_{q q''}(x) D^J_{q''q'}(y) d\mu(q''),\quad
D^J_{qq'}(e) = \delta(q,q').
\end{equation}

The matrix elements of the local representation $T^J$ satisfy a system of differential equations derived from the lift condition \eqref{loc_rep} and the skew-Hermiticity of the $\lambda$-representation:
\begin{equation}
\label{SysEqD}
\eta_i(x,\p_x) \D^J_{qq'}(x) = \ell_i(q,\p_q; J) \D^J_{qq'}(x),\quad
 \xi_i(x,\p_x) \D^J_{qq'}(x) = - \overline{\ell_i(q',\p_{q'}; J)} \D^J_{qq'}(x),
\end{equation}
where $i = 1, \dots, n$.
Here, $\xi_i = \xi_i(x,\p_x)$ and $\eta_i = \eta_i(x,\p_x)$ are the basis left- and right-invariant vector fields on $G$, respectively, regarded as first-order differential operators in the neighborhood $U_e$ of the group identity. 
Together with the initial condition $D^J_{qq'}(e) = \delta(q,q')$, this system~\eqref{SysEqD} uniquely defines the kernel in a neighborhood $U_e$ of the identity.

While formula~\eqref{loc_rep} defines the representation $T^J$ only locally on the Lie group, it can sometimes be extended to a global representation of the entire group~(see, for example, \cite{Kir76}).
A key obstruction to such a global extension is the requirement that the matrix elements be single-valued functions on~$G$.
From Eq.~\eqref{SysEqD}, this single-valuedness condition imposes a quantization constraint on the regular dual elements that define the $\lambda$-representations~\cite{Kir76, Shi00}:
\begin{equation}
\label{QuantCond}
\Gamma_\lambda([\gamma]) \defeq
\frac{1}{2\pi} \oint \limits_{\gamma \in [\gamma]} \omega^{\lambda} \in \mathbb{Z},\quad
[\gamma] \in H_1(G_\lambda).
\end{equation}
Here, $G_\lambda$ is the stabilizer of a covector $\lambda \in \g^*$ under the coadjoint action, $H_1(G_\lambda)$ is the one-dimensional homology group of $G_\lambda$, $\omega^\lambda$ is the left-invariant 1-form on $G_\lambda$ given by $\omega^\lambda(x) = (dL_x)^* \lambda$ for all $x \in G_\lambda$.
Consequently, the representation $T^J$ is globally defined on 
$G$ if and only if the corresponding dual element satisfies the quantization condition~\eqref{QuantCond}. 
This condition ensures the single-valuedness of the matrix elements $\D^J_{qq'}(x)$ on $G$ and selects a set of admissible parameters.
In other words, global representations correspond not to the entire parameter space $J$, but to the subset:
\begin{equation*}
\Lambda = \{ J \in \mathcal{J} \colon \Gamma_{\sigma(J)}([\gamma]) \in \mathbb{Z},\ [\gamma] \in H_1(G_{\sigma(J)}) \} \subset \mathcal{J}.
\end{equation*}

To simplify the presentation and avoid technical complications, we assume the Lie group~$G$ is unimodular. 
The subsequent results can be extended to the non-unimodular case with appropriate modifications (see, for example,~\cite{Bre10}).
For clarity, we will later illustrate these modifications in a detailed example involving a non-unimodular group.

Assuming the orthogonality and completeness of the matrix elements $\D^J_{qq'}(x)$, as expressed by the relations
\begin{equation}
\label{OrtConds}
\int_{G} \overline{\D^J_{qq'}(x)}\, \D^{\tilde{J}}_{\tilde{q}\tilde{q}'}(x)\, d\mu(x) = \delta(q,\tilde{q}) \delta(q',\tilde{q}') \delta(J,\tilde{J}),
\end{equation}
\begin{equation}
\label{ComConds}
\int_{Q^2 \times \Lambda} \overline{\D^J_{qq'}(x)}\, \D^{J}_{qq'}(\tilde{x})\, d\mu(q) d\mu(q') d\mu(J) = \delta(x,\tilde{x}),
\end{equation}
we can define a generalized Fourier transform by
\begin{equation}
\label{dirGFT}
\hat{\psi}(q, q'; J) =
\F[\psi](q, q'; J) =
\int_G \overline{\D^J_{q'q}(x)}\, \psi(x) d\mu(x),\quad
\psi \in L^2(G,d\mu(x)).
\end{equation}
Here, $d\mu(x)$ is a Haar measure on the unimodular Lie group $G$, $d\mu(J)$  is the spectral measure on the parameter set 
$\Lambda$ induced by the Casimir operators.  
The corresponding inverse transform is given by
\begin{equation}
\label{invGFT}
\psi(x) = 
\F^{-1}[\hat{\psi}](x) = 
\int_{Q^2 \times \Lambda} \D^J_{q'q}(x) \hat{\psi}(q, q'; J) d\mu(q) d\mu(q') d\mu(J).
\end{equation}
Furthermore, it follows directly from the system of equations \eqref{SysEqD} that the generalized Fourier transform intertwines the action of the Lie algebra. 
Specifically, the left- and right-invariant vector fields on 
$G$ map to the corresponding $\lambda$-representation operators:
\begin{equation}
\label{InterWinProp}
\F \circ \xi_i(x,\p_x) = \ell_i(q,\p_q; J) \circ \F,\quad
\F \circ \eta_i(x,\p_x) = -\overline{\ell_i(q', \p_{q'}; J)} \circ \F,\quad
i = 1, \dots, n.
\end{equation}

Note that while general theorems establishing the orthogonality and completeness of the matrix elements $\D^J_{qq'}(x)$ for an arbitrary Lie group is lacking, these properties are nevertheless expected to hold for the families of representations considered here. 
The standard approach in such cases is to verify these relations directly for each concrete realization of a group $G$ and its representations. 
In Section~\ref{ExaSec}, we demonstrate this procedure for two specific Lie groups.

The general formalism developed above provides a direct method for reducing the left-invariant Laplace--Beltrami equation~\eqref{LBeq}. 
Recall that these operators are constructed from the left-invariant vector fields $\xi_i$ (see Eq.~\eqref{xiLaplacian}).

Applying the generalized Fourier transform to Eq.~\eqref{LBeq} and using the intertwining property~\eqref{InterWinProp}, we obtain the \textit{reduced Laplace--Beltrami equation} on the representation space:
\begin{equation}
\label{redLBeq}
\redDelta \hat{\psi}(q,q'; J) = E \hat{\psi}(q,q'; J).
\end{equation}
Here, $\redDelta$ is the image of the original Laplace--Beltrami operator under the generalized Fourier transform
\begin{equation*}
\redDelta \defeq 
\mathcal{F} \circ \Delta \circ \mathcal{F}^{-1} = 
\sum_{i,j = 1}^n \G^{ij} \ell_i(q,\p_q; J) \ell_j(q, \p_q; J).
\end{equation*}
Note that the first-order terms from~\eqref{xiLaplacian} vanish in this expression due to the unimodularity of $G$, which simplifies the operator~$\Delta$.

In general, the reduced Laplace--Beltrami equation~\eqref{redLBeq} is not integrable. 
Nevertheless, several important special cases admit a complete solution.

The simplest such case occurs for a \textit{bi-invariant metric}---that is, a left-invariant metric which is also invariant under right translations. 
One can show that the left-invariant metric~\eqref{LI_metric} is bi-invariant if and only if the corresponding symmetric bilinear form $\G$ on~$\g$ is invariant under the adjoint representation:
\begin{equation*}
\G(\Ad(z) X, \Ad(z) Y) = \G(X,Y),\quad
X,Y \in \g,\quad
z \in G.
\end{equation*}
For such a metric, the Laplacian belongs to the center of the universal enveloping algebra $U(\gL)$, making it a second-order Casimir operator. 
Since the $\lambda$-representations are operator-irreducible, the generalized Fourier transform of this Casimir is a scalar:	
\begin{equation*}
\redDelta = \varkappa(J) \cdot \id.
\end{equation*}
Consequently, the reduced equation~\eqref{redLBeq} becomes purely algebraic and is trivially solvable.

Another class of exactly solvable cases arises when $\dim Q = 1$, or equivalently (by Eq.~\eqref{dimQ}), when the index of the algebra is $\dim \g - 2$. 
For these groups, the reduced Laplace--Beltrami equation reduces to a linear homogeneous second-order ODE, whose general solution can often be expressed in terms of special functions.

These two cases, however, by no means exhaust the possibilities. The power of the noncommutative reduction method lies in its ability to reveal hidden integrability beyond these traditional cases. 
In Section~\ref{sec:3}, we identify a new and substantially broader class of left-invariant metrics for which the reduced equation simplifies further---not merely to an ODE, but to a first-order PDE. 
This leads to explicit solutions and, remarkably, gives rise to nonlocal symmetry operators, demonstrating that the condition for integrability is far less restrictive than previously known examples might suggest.

\section{Left-invariant Laplace--Beltrami equations reducing to first-order PDEs}
\label{sec:3}

As can be seen from the previous section, a central problem is the integration of the reduced Laplace--Beltrami equation \eqref{redLBeq}, which is typically a second-order PDE. This section identifies a special class of left-invariant pseudo-Riemannian metrics that simplify this problem considerably, as their corresponding reduced equations are of first order.

Let $(\g, \G)$ be an $n$-dimensional pseudo-Riemannian Lie algebra.
Our central hypothesis is that $\g$ contains a commutative ideal $\h$ whose orthogonal complement $\h^\perp$ with respect to~$\G$ is contained within itself:
\begin{equation}
\label{hperp}
\h^\perp \subseteq \h.   
\end{equation}  
A commutative ideal satisfying this condition will be called a \textit{coisotropic ideal} in the pseudo-Riemannian Lie algebra $(\g,\G)$.
 
Let us analyze the consequences of condition~\eqref{hperp}. 
First, we derive constraints on the dimension of the ideal~$\h^\perp$. 
From the standard identity $\dim \h + \dim \h^\perp = \dim \g$ and condition~\eqref{hperp}, it follows immediately that
\begin{equation*}
\dim \h \geq \frac{1}{2}\, \dim \g.
\end{equation*}
Furthermore, since $\h^\perp$ is contained in $\h$, the bilinear form $\G$ vanishes identically on $\h^\perp$, making it a null subspace.
This leads to a stronger restriction.
The maximal dimension of a null subspace in an $n$-dimensional space with signature $(p,n-p)$ is $\min \{ p, n-p \}$~\cite{Cec92}.
Therefore, we have the inequality:
\begin{equation*}
\label{ineq_dimh}
\dim \h \geq \dim \g - \min \, \{ p, n - p \}.
\end{equation*}
This restriction has immediate consequences for specific signatures:
\begin{itemize}
\item In the Riemannian case (signature $(n,0)$), we find $\dim \h \geq \dim \g$, forcing $\g = \h$. Hence, the Lie algebra $\g$ must be commutative.
\item In the Lorentzian case (signature $(n-1,1)$), the restriction becomes $\dim \h \geq \dim \g - 1$. This permits the non-trivial case where $\h$ is a commutative ideal of codimension one.
\end{itemize}

In the Lorentzian case, the condition $\mathfrak{h}^{\perp}\subseteq\mathfrak{h}$ implies $\dim\mathfrak{h} = \dim\mathfrak{g} - 1$; that is, $\mathfrak{h}$ is a commutative ideal of codimension one. 
Lie algebras with this property are called \textit{almost abelian} and have been extensively studied in the literature (see, e.g., \cite{Gor2020} and references therein). 
In particular, all such algebras are solvable and admit an explicit matrix realization. 
For signatures of higher index (e.g., $(2,2)$, as in Example~\ref{sec:4-2} below), our construction leads to a broader class of metrics, including non-unimodular algebras that are not almost abelian in the classical sense. 
Thus, our approach naturally generalizes known results on almost abelian Lie groups.

Since the form $\G$ is nondegenerate, it induces an isomorphism $\G^\flat \colon \g \to \g^*$ given by $\< \G^\flat(X), Y \> \defeq \G(X,Y)$ for $X,Y \in \g$.
Let $\h^0$ be the annihilator of the ideal $\h$ in the dual space $\g^*$, i.~e.,
\begin{equation*}
\h^0 = \{ f \in \g^* \colon \< f, X \> = 0,\ X \in \h \}.
\end{equation*}
A key observation is that the restriction of the isomorphism $\G^\flat$ to the orthogonal complement $\h^\perp$ is an isomorphism between $\h^\perp$ and $\h^0$. 
Consequently, the inclusion from~\eqref{hperp} is equivalent to the condition that the inverse isomorphism $\G^\sharp \colon \g^* \to \g$ maps the annihilator back into the ideal:
\begin{equation}
\label{Gf_in_h}
\G^\sharp(\h^0) \subseteq \h.
\end{equation}

Let $\e = \{ e_1, \dots, e_n \}$ be a basis of the Lie algebra $\g$ that is adapted to the ideal $\h$; that is, the first vectors $e_1, \dots, e_s$ form a basis of~$\h$, $s = \dim \h$.
Let $\e^* = \{ e^1, \dots, e^n \}$ be the basis of $\g^*$ dual to $\e = \{ e_1, \dots, e_n \}$, so that $\< e^i, e_j \> = \delta^i_j$ for $i,j = 1, \dots, n$. 
In this basis, the annihilator $\h^0$ is spanned by the $\{ e^{s+1}, \dots, e^n \}$. 
The components $\G_{ij} = \G(e_i, e_j)$ are the matrix elements of the isomorphism $\G^\flat \colon \g \to \g^*$ relative to the bases $\e$ and $\e^*$, so that $\G^\flat(e_j) = \sum_{i = 1}^n \G_{ij} e^i$ for $j = 1, \dots, n$.
Let $\G^{ij}$ be the entries of the inverse matrix, representing the map $\G^\sharp \colon \g^* \to \g$.
The condition~\eqref{Gf_in_h} --- that $\G^\sharp$ maps $\h^0$  into $\h$ --- is equivalent to the vanishing of the bottom-right $(n-s)\times(n-s)$ block of this matrix:
\begin{equation*}
\G^{ab} = 0,\quad
a,b = s + 1, \dots, n.
\end{equation*}
This block structure has a direct consequence for the Laplace--Beltrami operator, whose general form is given by~\eqref{xiLaplacian}.
The condition $\G^{ab} = 0$ for $a,b \geq s + 1$ eliminates all
terms quadratic in the vector fields $\xi_{s+1}, \dots, \xi_n$
that lie outside the commutative ideal~$\h$. 
Consequently, the operator $\Delta$ becomes \textit{linear} in these fields.
Performing the sum while excluding the vanishing terms yields the explicit form:
\begin{equation}
\label{1ord_Delta}
\Delta = 
\sum_{\alpha = 1}^s \sum_{\beta = 1}^s \G^{\alpha\beta} \xi_\alpha \xi_\beta +  2 \sum_{\alpha = 1}^s \sum_{b = s + 1}^n \G^{\alpha b} \xi_\alpha \xi_b + \sum_{\alpha = 1}^s \mathbf{B}^\alpha \xi_\alpha,
\end{equation}
where the constants $\mathbf{B}^\alpha$ are equal to $\mathbf{B}^\alpha \defeq \C^\alpha - \sum_{\beta = 1}^s \sum_{b = s + 1}^n C_{\beta b}^\alpha \G^{\beta b}$ for $\alpha = 1, \dots, s$. 

For the class of Lie algebras satisfying condition~\eqref{hperp}, one can show that for any regular $\lambda \in \g^*$ there exists a polarization $\pol \in \mathfrak{P}(\lambda)$ with $\h \subseteq \pol$. 
This fact, whose proof is rather technical, guarantees that the $\lambda$-representation operators corresponding to $\h$ act as multiplication.
Recall that $\pol$ is the isotropy subalgebra of the homogeneous space $Q = G/P$ on which $\lambda$-representations of the Lie algebra $\g$ are realized (see Sec.~\ref{sec:2}).
The inclusion $\h \subset \pol$ implies that the generators of the action of $G$ on $Q$ associated with the ideal $\h$ vanish identically.
Consequently, in the $\lambda$-representation, the operators $\ell_\alpha$ corresponding to the basis elements $e_\alpha \in \h$ act as multiplication operators:
\begin{equation}
\label{l_alpha}
\ell_\alpha(q,\p_q; J) = i \chi_\alpha(q; J),\quad
\alpha = 1, \dots, s.
\end{equation}
This observation has a profound simplifying effect.
When the generalized Fourier transform $\mathcal{F}$ is applied, the Laplace--Beltrami equation with the Laplacian~\eqref{1ord_Delta} reduces to a first-order linear PDE. 
This follows directly by substituting~\eqref{l_alpha} into the expression for the reduced operator $\redDelta$, derived from \eqref{1ord_Delta}. The result is:
\begin{equation}
\label{FirstOredrRedLaplace}
\redDelta =
2 i \sum_{A = 1}^m Z^A(q; J) \frac{\p}{\p q^A} + V(q; J) ,
\end{equation}
where $m = \dim Q$, and the coefficient functions $Z^A(q; J)$ and $V(q; J)$ are given by:
\begin{align*}
Z^A(q; J) &= \sum_{\alpha = 1}^s \sum_{b = s + 1}^n \G^{\alpha b} \chi_\alpha(q; J) \zeta_b^A(q),\quad
A = 1, \dots, m,\\
V(q; J) &= - \sum_{\alpha = 1}^s \left ( \sum_{\beta = 1}^s \G^{\alpha\beta} \chi_\beta(q;J) + 2 \sum_{b = s + 1}^n \G^{\alpha b} \chi_b(q; J) + i \mathbf{B}^\alpha \right ) \chi_\alpha(q; J).
\end{align*}
Recall from Sec.~\ref{sec:2} that $\ell_b(q,\p_q; J) = \zeta_b^A(q) \p_{q^A} + \chi_b(q; J)$, where $\zeta_b$ are the generators of the $G$-action on $Q$

Let $q_0 \in Q$ be a point where the vector field $Z = \sum_{A = 1}^m Z^A(q; J) \p/\p q^A$ is non-vanishing.
By the straightening (or flow box) theorem for vector fields (see, for example, \cite[p.~30]{Olv93}), there exists a local coordinate chart $(u^1, \dots, u^{m - 1}, v)$ defined in a neighborhood of $q_0$ such that $Z = \p/\p v$.
Since we are ultimately interested in constructing local solutions to the original equation, the local nature of this coordinate transformation is sufficient for our purposes.
In these straightened coordinates, the reduced Laplace--Beltrami operator \eqref{FirstOredrRedLaplace} simplifies to:
\begin{equation}
\label{redLBop}
\redDelta =
2 i \frac{\p}{\p v} + V(q(u,v); J),
\end{equation}
where $q^A = q^A(u, v)$ represents the coordinate transformation from $(u,v) = (u^1, \dots, u^{m - 1}, v)$ to $q = (q^1, \dots, q^{m})$.
Now, define the function $\hat{\varphi}(u,v,q'; J) \defeq \hat{\psi}(q(u,v),q'; J)$. 
Substituting this and the simplified operator~\eqref{redLBop} into the reduced equation yields a first-order PDE:
\begin{equation*}
\frac{\p \hat{\varphi}(u,v,q'; J)}{\p v} = - \frac{i}{2} \left [ E - V(q(u,v); J) \right ] \hat{\varphi}(u,v,q'; J).
\end{equation*}
This equation can be integrated directly, leading to the general solution:
\begin{equation*}
\hat{\varphi}(u,v,q'; J) = \Phi(u, q'; J) \exp \left ( -\frac{i}{2} \int \left [ E - V(q(u,v); J) \right ] dv \right ).
\end{equation*}
Here, $\Phi(u, q'; J)$ is an arbitrary function of the variables $u$, $q'$, and the spectral parameter $J$, which play the role of the constants of integration for this ordinary differential equation in~$v$.

We conclude by noting that the left-invariant Laplace--Beltrami  equations studied here can, in some cases, possess \textit{accidental symmetries} --- symmetries not generated by the natural geometric action of right-invariant vector fields.
The origin of these symmetries lies in the structure of the reduced operator~$\redDelta$.
Recall that the coordinates $(u^1, \dots, u^{m-1})$ (where $ m = \dim Q$) are functionally independent invariants of the flow generated by the vector field $Z$; they can be obtained as first integrals of the system of ordinary differential equations $dq^A/dt = Z^A(q)$. 
Consequently, they commute with the reduced operator:
\begin{equation*}
[\redDelta, u^A] = 0,\quad
A = 1, \dots, m - 1,
\end{equation*}
where each $u^A(q)$ is viewed as a multiplication operator. This establishes the $u^A$ as symmetry operators for the reduced Laplace--Beltrami equation.
Lifting these symmetries back to the group via the inverse generalized Fourier transform yields symmetry operators for the original Laplace--Beltrami equation on $G$:
\begin{equation*}
Y_A = \mathcal{F}^{-1} \circ u^A(q) \circ \mathcal{F},\quad
A = 1, \dots, m - 1.
\end{equation*}
A remarkable feature of these operators $Y_A$ is that they are generally not standard differential operators. 
Instead, they are often integro-differential operators, reflecting the non-local character of the transform~$\mathcal{F}$. 
This nonlocality arises because the inverse generalized Fourier transform $\mathcal{F}^{-1}$ intertwines multiplication operators on $Q$ with nonlocal convolution-type operators on $G$.
This important phenomenon will be illustrated by the second example in the following section. 

\section{Examples}
\label{ExaSec}
\subsection{Heisenberg group}

As a first example, we consider the three-dimensional Heisenberg group $G = \H{3}{\R}$.
Its relative simplicity provides a clear setting to highlight the distinctive features of noncommutative harmonic analysis, contrasting it with the commutative framework of separation of variables.

The group $G$ is realized as the group of $3 \times 3$ upper triangular matrices of the form
\begin{equation*}
h(x) = 
\left (
\begin{array}{ccc}
1	&	x_1	&	x_3	\\
0	&	1	&	x_2	\\
0	&	0	&	1
\end{array}
\right ),\quad
\text{where }
x = (x_1, x_2, x_3) \in \R^3.
\end{equation*}
The parameters $x_1, x_2, x_3$ serve as global coordinates on $G$,  in which the group multiplication is given by
In these coordinates, the group multiplication law $h(z) = h(x) h(y)$ is given by
\begin{equation}
\label{HeisMultLow}
z_1 = x_1 + y_1,\quad
z_2 = x_2 + y_2,\quad
z_3 = x_3 + y_3 + x_1 y_2.
\end{equation}

The Lie algebra $\g$ of the Heisenberg group consists of the $3 \times 3$ strictly upper triangular matrices, i.~e., those with zeros on and below the main diagonal. 
A natural basis for $\g$ is given by the matrices:
\begin{equation*}
e_1 = 
\left (
\begin{array}{ccc}
0	&	1	&	0	\\
0	&	0	&	0	\\
0	&	0	&	0
\end{array}
\right ),\quad
e_2 = 
\left (
\begin{array}{ccc}
0	&	0	&	0	\\
0	&	0	&	1	\\
0	&	0	&	0
\end{array}
\right ),\quad
e_3 = 
\left (
\begin{array}{ccc}
0	&	0	&	1	\\
0	&	0	&	0	\\
0	&	0	&	0
\end{array}
\right ).
\end{equation*}
These basis elements satisfy the commutation relations:
\begin{equation*}
[e_1, e_2]_\g = e_3,\quad
[e_1, e_3]_\g = 
[e_2, e_3]_\g = 0.
\end{equation*}
This structure shows that $\g$ is a central extension of the abelian algebra $\langle e_1, e_2 \rangle \simeq \R^2$  by the one-dimensional center $\langle e_3 \rangle \simeq \R$.
The index of this Lie algebra is equal to one: $\ind \g = 1$.

The group multiplication law~\eqref{HeisMultLow}yields the following left- and right-invariant vector fields:
\begin{equation*}
\xi_1 = \frac{\p}{\p x_1},\quad
\xi_2 = \frac{\p}{\p x_2} + x_1 \, \frac{\p}{\p x_3},\quad
\xi_3 = \frac{\p}{\p x_3};
\end{equation*}
\begin{equation}
\label{HeisRightVF}
\eta_1 = \frac{\p}{\p x_1} + x_2 \, \frac{\p}{\p x_3},\quad
\eta_2 = \frac{\p}{\p x_2},\quad
\eta_3 = \frac{\p}{\p x_3}.
\end{equation}
The dual 1-forms $\omega^i$ to the left-invariant fields $\xi_i$ are given by
\begin{equation*}
\omega^1 = dx_1,\quad
\omega^2 = dx_2,\quad
\omega^3 = - x_1 dx_2 + dx_3.
\end{equation*}
The left- and right-invariant Haar measures coincide,
\begin{equation*}
d\mu(x) \defeq d\mu_L(x) = d\mu_R(x) = dx_1 \wedge dx_2 \wedge dx_3.
\end{equation*}
so the Heisenberg group is unimodular.

The classification of left-invariant metrics on the three-dimensional Heisenberg group is well-known~\cite{Rah92, Vuk15}.
Up to automorphism of $\g$ and homothety, there is a unique left-invariant Riemannian metric, but Rahmani~\cite{Rah92} showed there are three distinct Lorentzian metrics. 
These are classified by the nature of the center of the Lie algebra $\g$ with respect to the induced symmetric bilinear form $\G$.
The metric with a null center corresponds to the case studied in Section~\ref{sec:3}.
To see this, consider the commutative ideal $\h = \langle e_1, e_3 \rangle$.
With respect to the bilinear form
\begin{equation*}
\G = e^1 \otimes e^1 + e^2 \otimes e^3 + e^3 \otimes e^2,
\end{equation*}
the orthogonal subspace is $\h^\perp = \langle e_3 \rangle$,
which satisfies $\h^\perp \subset \h$, confirming that the center is null.
In the global coordinates $x_1, x_2, x_3$, this metric is expressed in terms of the left-invariant 1-forms as
\begin{equation}
\label{HeisMetric}
g = 
\omega^1 \otimes \omega^1 + \omega^2 \otimes \omega^3 + \omega^3 \otimes \omega^2 = 
dx_1^2 - 2 x_1 dx_2^2 + 2 dx_2 dx_3.
\end{equation}
The corresponding Laplace--Beltrami equation is then
\begin{equation}
\label{HeisDelta}
\Delta \psi = 
\left ( \xi_1^2 + 2 \xi_2 \xi_3 \right ) \psi = 
\frac{\p^2 \psi}{\p x_1^2} + 2\, \frac{\p^2 \psi}{\p x_2 \p x_3} + 2 x_1 \, \frac{\p^2 \psi}{\p x_3^2} = E \psi.
\end{equation}

Although the Lorentzian metric~\eqref{HeisMetric} is flat and thus possesses the full six-dimensional isometry group $\mathrm{IO}(1,2)$ of $\R^{1,2}$, we will restrict our analysis to the three-dimensional subalgebra of right-invariant vector fields $\gR = \langle \eta_1, \eta_2, \eta_3 \rangle$.
This is sufficient to solve the Laplace--Beltrami equation~\eqref{HeisDelta}, and the additional Killing vectors are not required for this purpose.

Before applying noncommutative reduction, we first examine a commutative approach to solving the Laplace–Beltrami equation.
The right-invariant vector fields $\eta_2 = \p/\p x_2$ and $\eta_3 = \p/\p x_3$ commute with the Laplacian $\Delta$, allowing us to seek solutions of Eq.~\eqref{HeisDelta} in the form
\begin{equation*}
\psi_{\mu, \nu}(x_1, x_2, x_3) = 
\exp \left ( i \mu x_2 + i \nu x_3 \right ) \varphi_{\mu,\nu}(x_1),
\end{equation*}
where $\varphi_{\mu, \nu}(x_1)$ is a function to be determined, and $\mu, \nu \in \mathbb{R}$ with $\nu \neq 0$.
Substituting this ansatz into~\eqref{HeisDelta} reduces it to the second-order ODE
\begin{equation*}
\varphi''_{\mu,\nu}(x_1) - \left ( 2 \nu^2 x_1 + 2 \nu \mu + E \right ) \varphi_{\mu,\nu}(x_1) = 0.
\end{equation*}
The general solution is given in terms of Airy functions:
\begin{equation*}
\varphi_{\mu, \nu}(x_1) = 
A(\mu, \nu) \cdot \Ai \left ( \frac{2 \nu^2 x_1 + 2 \mu \nu + E}{( 2 \nu^2 )^{2/3}} \right ) + 
B(\mu, \nu) \cdot \Bi \left ( \frac{2 \nu^2 x_1 + 2 \mu \nu + E}{( 2 \nu^2 )^{2/3}} \right ),
\end{equation*}
where $A$ and $B$ are arbitrary functions of the parameters $\mu$ and $\nu$.
Imposing the boundary condition $\psi \to 0$ as $x_1 \to \infty$ requires $B(\mu,\nu) = 0$, as is unbounded.
The resulting general solution of Eq.~\eqref{HeisDelta} is then the superposition
\begin{equation}
\label{HeisGenSol}
\psi(x_1, x_2, x_3) = 
\int_{-\infty}^{+\infty} d \mu  
\int_{-\infty}^{+\infty} d \nu \,  
A( \mu, \nu ) \Ai \left ( \frac{2 \nu^2 x_1 + 2 \mu \nu + E}{( 2 \nu^2 )^{2/3}} \right ) \exp \left ( i \nu x_2 + i \mu x_3 \right ).
\end{equation}

We now apply the noncommutative reduction method from Section~\ref{sec:2} to construct solutions of the Laplace--Beltrami equation~\eqref{HeisDelta}.

The regular coadjoint orbits in $\g^*$ are the planes $\mathcal{O}_J = \{ f_1 e^1 + f_2 e^2 + J e^3 \colon f_1, f_2 \in \R \}$, parameterized by $J \in \mathcal{J} = \R \setminus \{ 0 \}$.
We fix a section $\sigma \colon \mathcal{J} \to \g^*$ by choosing the representative $\sigma(J) = J e^3$ for each orbit $\mathcal{O}_J$
The commutative ideal $\h = \langle e_1, e_3 \rangle$ is a polarization at every point $\sigma(J)$.
The corresponding $\lambda$-representation, which acts on the space $L^2(\R, dq)$ with $q \in Q = \H{3}{\R} / \R^2 \simeq \R$, is given by the skew-symmetric operators:
\begin{equation*}
\ell_1 = - i J q,\quad
\ell_2 = \frac{\p}{\p q},\quad
\ell_3 = i J.
\end{equation*}
(For an efficient algebraic method of constructing $\lambda$-representations, see Shirokov~\cite{Shi00}.)
Solving system of first-order PDE's~\eqref{SysEqD} yields the matrix elements $\D^J_{qq'}(x)$ of the associated Lie group representation $T^J(x)$:
\begin{equation}
\label{HeisD}
\D^J_{qq'}(x) = \delta( q + x_2 - q' ) \exp \left ( - i J q' x_1 + i J x_3 \right ).
\end{equation}
These matrix elements satisfy the normalization condition $\D^J_{qq'}(e) = \delta(q - q')$, where $\delta$ is the usual Dirac delta-function on $\R$.
Consequently, the action of the representation $T^J$ on a function $\varphi \in L^2(\R, dq)$ is explicitly given by:
\begin{equation*}
T^J(x) \varphi(q) = 
\int_{-\infty}^{+\infty} \D^J_{qq'}(x) \varphi(q') dq' = 
\exp \left [ - i J ( q + x_2 ) x_1 + i J x_3 \right ] \varphi( q + x_2 ).
\end{equation*}

The matrix elements~\eqref{HeisD} satisfy the orthogonality and completeness relations:
\begin{equation*}
\int \limits_{\H{3}{\R}} \overline{\D^J_{qq'}(x)} \D^{\tilde{J}}_{\tilde{q}\tilde{q}'}(x) d\mu(x) = 
\delta \left ( q - \tilde{q} \right ) \delta \left ( q' - \tilde{q}' \right ) \delta(J, \tilde{J}),
\end{equation*}
\begin{equation*}
\int \limits_{\R^2 \times \mathcal{J}} \overline{\D^J_{qq'}(x)} \D^J_{q q'}(\tilde{x}) dq dq' d\mu(J) = 
\delta( x_1 - \tilde{x}_1 ) \delta ( x_2 - \tilde{x}_2 ) \delta ( x_3 - \tilde{x}_3 ),
\end{equation*}
where the measure $d\mu(J)$ and the delta function $\delta(J, \tilde{J})$ on the set $\mathcal{J}$ are given by
\begin{equation*}
d\mu(J) = \frac{|J| dJ}{(2\pi)^2},\quad
\delta(J, \tilde{J}) = \frac{(2\pi)^2}{|J|}\, \delta(J - \tilde{J}).
\end{equation*}
These relations allow us to define direct and inverse generalized Fourier transforms, $\F$ and $\F^{-1}$, on the Heisenberg group (see Eqs.~\eqref{dirGFT} and \eqref{invGFT}).
Applying the direct transform $\F$ to the Laplace--Beltrami equation~\eqref{HeisDelta} reduces it to the differential equation
\begin{equation*}
\tilde{\Delta} \hat{\psi}(q,q'; J) = 
\left ( \ell_1^2 + 2 \ell_2 \ell_3 \right ) \hat{\psi}(q,q'; J) = E \hat{\psi}(q,q'; J).
\end{equation*}
Substituting the explicit form of the operators $\ell_i$ yields a first-order ordinary differential equation in $q$:
\begin{equation*}
2 i J \, \frac{\p \hat{\psi}(q,q'; J)}{\p q} - J^2 q^2 \hat{\psi}(q,q'; J) = E \hat{\psi}(q,q'; J).
\end{equation*}
The general solution to this equation is
\begin{equation}
\label{HeisHatPsi}
\hat{\psi}(q,q'; J) = \Phi(q'; J) \exp \left ( - \frac{i J q^3}{6} - \frac{i E q}{2 J} \right ),
\end{equation}
where $\Phi(q'; J)$ is an arbitrary function of the parameters $q'$ and $J$.

Let $\hat{\Phi}(k; J)$ denote the Fourier transform of $\Phi(q'; J)$ with respect to the variable~$q'$:
\begin{equation}
\label{HeisPhiFT}
\Phi(q'; J) = \frac{1}{2\pi} \int_{-\infty}^{+\infty} \hat{\Phi}(k; J) e^{-i k q'} \, dk.
\end{equation}
The general solution to the Laplace--Beltrami equation~\eqref{HeisDelta} is obtained by substituting expressions~\eqref{HeisHatPsi} and~\eqref{HeisPhiFT} into the inverse generalized Fourier transform~\eqref{invGFT}.
Employing the integral identity for the Airy function~\cite{Wid79},
\begin{equation*}
\frac{1}{2\pi} \int_{-\infty}^{+\infty} e^{i x \tau + i t \tau^3} \, d\tau = 
\frac{1}{(3|t|)^{1/3}}\, \Ai \left ( \frac{\mathrm{sign}(t) x}{(3|t|)^{1/3}} \right ), 
\end{equation*} 
we derive the final result:
\begin{multline*}
\psi(x_1, x_2, x_3) = 
\\
= \frac{1}{(2\pi)^2} 
\int_{-\infty}^{+\infty} dk 
\int_{-\infty}^{+\infty} dJ \, 
\left ( 2 |J|^2 \right )^{1/3} \hat{\Phi}(k; J) \mathrm{Ai} \left ( \frac{2 J^2 x_1 + 2 k J + E}{(2J^2)^{2/3}} \right ) \exp(i k x_2 + i J x_3).
\end{multline*}
A direct comparison shows that this expression is equivalent to~\eqref{HeisGenSol} under the identification
\begin{equation*}
k = \mu,\quad
J = \nu,\quad
\frac{\left ( 2 |J|^2 \right )^{1/3}}{(2\pi)^2}\, \hat{\Phi}(k; J) = A(\mu,\nu).
\end{equation*}

This example successfully demonstrates the application of noncommutative harmonic analysis to solving PDEs on the Heisenberg group. 
The method, utilizing generalized Fourier transforms and $\lambda$-representations, provides an alternative to the classical separation of variables. 
Crucially, it reproduces the general solution obtained by the commutative approach, thereby validating the consistency of the noncommutative framework.

\subsection{Four-dimensional non-unimodular Lie group}
\label{sec:4-2}

We now proceed to a more nontrivial example where the proposed class of left-invariant pseudo-Riemannian metrics gives rise to non-obvious symmetries of the corresponding Laplace--Beltrami operator.

We consider the four-dimensional Lie algebra $\g$ with the non-vanishing commutation relations
\begin{equation*}
[e_1, e_4]_\g = 2 e_1,\quad
[e_2, e_3]_\g = e_1,\quad
[e_2, e_4]_\g = e_2,\quad
[e_3, e_4]_\g = e_2 + e_3.
\end{equation*}
This indecomposable algebra is Frobenius, as its index is zero.
In the Mu\-ba\-rak\-zyanov classification~\cite{Mub63}, it is labeled as~$\g_{4,7}$.
The associated simply connected Lie group $G$ admits a global parametrization by $x = (x_1, x_2, x_3, x_4) \in \R^4$ via the matrix representation~\cite{BigRem16}:
\begin{equation*}
h(x) = 
\left (
\begin{array}{rrrr}
e^{-2x_4}	&	- x_3 e^{-x_4}	&	\left ( x_2 + x_3 x_4 \right )e^{-x_4}	&	2 x_1	\\
0	&	e^{-x_4}	&	- x_4 e^{-x_4}	&	x_2	\\
0	&	0	&	e^{-x_4}	&	x_3	\\
0	&	0	&	0	&	1
\end{array}
\right ).
\end{equation*}
The group multiplication $h(x)h(y)=h(z)$ is given by
\begin{align*}
z_1 &= x_1 + e^{-2x_4} y_1 + \frac{1}{2}\, e^{-x_4} \left ( x_2 y_3 - x_3 y_2 + x_3 x_4 y_3 \right ),\\
z_2 &= x_2 + e^{-x_4} \left ( y_2 - x_4 y_3 \right ),\\
z_3 &= x_3 + e^{-x_4} y_3,\\
z_4 &= x_4 + y_4.
\end{align*}
The left- and right-invariant vector fields on $G$ are, respectively,
\begin{equation*}
\xi_1 = e^{-2x_4} \frac{\p}{\p x_1},\quad
\xi_2 = e^{-x_4} \left ( \frac{\p}{\p x_2} - \frac{x_3}{2}\,\frac{\p}{\p x_1} \right ),\quad
\xi_3 = e^{-x_4} \left ( \frac{\p}{\p x_3} - x_4 \frac{\p}{\p x_2} + \frac{x_2 + x_3 x_4}{2}\, \frac{\p}{\p x_1} \right ),
\end{equation*}
\begin{equation*}
\xi_4 = \frac{\p}{\p x_4},\quad
\eta_1 = \frac{\p}{\p x_1},\quad
\eta_2 = \frac{\p}{\p x_2} + \frac{x_3}{2}\, \frac{\p}{\p x_1},\quad
\eta_3 = \frac{\p}{\p x_3} - \frac{x_2}{2}\, \frac{\p}{\p x_1},
\end{equation*}
\begin{equation*}
\eta_4 = - 2 x_1 \frac{\p}{\p x_1} - \left ( x_2 + x_3 \right ) \frac{\p}{\p x_2} - x_3 \frac{\p}{\p x_3} + \frac{\p}{\p x_4}.
\end{equation*}
The left-invariant 1-forms dual to $\xi_i$ are
\begin{equation*}
\omega^1 = e^{2x_4} \left ( dx_1 + \frac{x_3 dx_2 - x_2 dx_3}{2} \right ),\quad
\omega^2 = e^{x_4} \left ( dx_2 + x_4 dx_3 \right ),\quad
\omega^3 = e^{x_4} dx_3,\quad
\omega^4 = dx_4.
\end{equation*}
The group $G$ is non-unimodular, as the left and right Haar measures differ:
\begin{equation*}
d\mu_L(x) = e^{4 x_4} dx_1 \wedge dx_2 \wedge dx_3 \wedge dx_4 \neq
d\mu_R(x) = dx_1 \wedge dx_2 \wedge dx_3 \wedge dx_4. 
\end{equation*} 

The Lie algebra $\g$ possesses a unique two-dimensional commutative ideal $\h = \langle e_1, e_2 \rangle$.
We consider non-degenerate symmetric bilinear forms on $\g$ satisfying the condition $\h^\perp \subseteq \h$.
Using automorphisms that preserve the ideal $\h$, any such form $\G$ can be brought into one of the following five canonical types:
\begin{align}
\nonumber
\G^{(1)} &= e^1 \otimes e^3 + e^3 \otimes e^1 + \alpha \left ( e^1 \otimes e^4 + e^4 \otimes e^1 \right ) + \beta \left ( e^2 \otimes e^4 + e^4 \otimes e^2 \right ),\quad \alpha \beta \neq 0,\\
\nonumber
\G^{(2)} &= e^1 \otimes e^3 + e^3 \otimes e^1 + \alpha \left ( e^2 \otimes e^4 + e^4 \otimes e^2 \right ) + \beta e^4 \otimes e^4,\quad \alpha \neq 0,\\
\nonumber
\G^{(3)} &= \varepsilon \left ( e^1 \otimes e^4 + e^4 \otimes e^1 \right ) + \alpha \left ( e^2 \otimes e^3 + e^3 \otimes e^2 \right ),\quad
\varepsilon = \pm 1,\ \alpha \neq 0,\varepsilon,\\
\nonumber
\G^{(4)} &= \varepsilon \left ( e^1 \otimes e^4 - e^2 \otimes e^3 - e^3 \otimes e^2 + e^4 \otimes e^1 \right ) + \alpha \left ( e^3 \otimes e^4 + e^4 \otimes e^3 \right ),\quad
\varepsilon = \pm 1,\\
\nonumber
\G^{(5)} &= \varepsilon \left ( e^1 \otimes e^4 + e^2 \otimes e^3 + e^3 \otimes e^2 + e^4 \otimes e^1 \right ) + \alpha \left ( e^2 \otimes e^4 + e^4 \otimes e^2 \right ),\quad
\varepsilon = \pm 1.
\end{align}
It can be shown that the left-invariant metrics associated with the forms $\G^{(2)} - \G^{(5)}$ admit an additional Killing vector field beyond the four right-invariant vector fields $\eta_i$. 
Since our primary interest lies in cases with minimal symmetry, we restrict our subsequent analysis to the metric defined by $\G = \G^{(1)}$, where the symmetry algebra is exactly the right-invariant vector fields $\gR$, with no additional Killing vectors.
In the global coordinates $x = (x_1, x_2, x_3, x_4)$, this left-invariant metric is given by
\begin{multline}
\label{Ex2Metrica}
g = 
\omega^1 \otimes \omega^3 + \omega^3 \otimes \omega^1 + 
\alpha \left ( \omega^1 \otimes \omega^4 + \omega^4 \otimes \omega^1 \right ) +
\beta \left ( \omega^2 \otimes \omega^4 + \omega^4 \otimes \omega^2 \right ) =
\\
= 2 e^{3x_4} dx_1 dx_3 + 2 \alpha e^{2x_4} dx_1 dx_4 + x_3 e^{3x_4} dx_2 dx_3 + \left ( \alpha x_3 e^{2x_4} + 2 \beta e^{x_4} \right ) dx_2 dx_4 -  
\\
- x_2 e^{3x_4} dx_3^2 + \left ( 2 \beta x_4 e^{x_4} - \alpha x_2 e^{2x_4} \right ) dx_3 dx_4.
\end{multline}
This defines a non-flat pseudo-Riemannian manifold of signature $(2,2)$.

According to Eq.~\eqref{xiLaplacian}, the Laplace–Beltrami operator defined by the metric~\eqref{Ex2Metrica} takes the following form in terms of left-invariant vector fields:
\begin{equation}
\label{Ex2LBEq}
\Delta \psi \defeq 
\left [ 2 \xi_1 \xi_3 + \frac{2}{\beta}\, \xi_2 \left ( \xi_4 - \alpha \xi_3 \right ) + \frac{1}{\beta} \left ( \alpha \xi_1 +3 \xi_2 \right ) \right ] \psi = E \psi.
\end{equation}
Its explicit coordinate expression is given by:
\begin{multline*}
\left ( e^{-3 x_4} + \frac{\alpha x_3}{2 \beta}\, e^{-2x_4} \right ) \left ( x_2 + x_3 x_4 \right ) \frac{\p^2 \psi}{\p x_1^2} - 
\left ( 2 x_4 e^{-3 x_4} + \frac{\alpha}{\beta} \left ( x_2 + 2 x_3 x_4 \right ) e^{-2x_4} \right ) \frac{\p^2 \psi}{\p x_1 \p x_2} + 
\\
+ \left ( 2 e^{-3x_4} + \frac{\alpha x_3}{\beta}\,e^{-2x_4}  \right ) \frac{\p^2 \psi}{\p x_1 \p x_3} - \frac{x_3}{\beta}\, e^{-x_4}\, \frac{\p^2 \psi}{\p x_1 \p x_4} + 
\frac{2 \alpha x_4}{\beta}\, e^{-2x_4}\, \frac{\p^2 \psi}{\p x_2^2} -
\\
- \frac{2\alpha}{\beta}\, e^{-2x_4} \frac{\p^2 \psi}{\p x_2 \p x_3} + \frac{2}{\beta}\, e^{-x_4} \, \frac{\p^2 \psi}{\p x_2 \p x_4} - \frac{3}{2\beta}\, x_3 e^{-x_4} \, \frac{\p \psi}{\p x_1} + \frac{3}{\beta}\, e^{-x_4}\, \frac{\p \psi}{\p x_2} = E \psi.
\end{multline*}
The absence of three-dimensional commutative subalgebras in $\g$ makes the classical separation of variables for Eq.~\eqref{Ex2LBEq} a non-trivial problem. 
A full solution would require the computation of second-order Killing tensors, which is a challenging task in itself~\cite{Sha79, Ben16}. This complexity motivates the application of the noncommutative integration method, which does not rely on a complete set of commutative symmetries.

We now construct exact solutions to Eq.~\eqref{Ex2LBEq} using the method described in Sec.~\ref{sec:2}.

The coadjoint representation admits precisely two regular orbits, which are the open connected sets
\begin{equation*}
\mathcal{O}_J = \{ f \in \g^* \colon J f_1 > 0 \},\quad
J = \pm 1.
\end{equation*}
We fix the representatives of these orbits as $\sigma(J) = J e^1 \in \g^*$.
The commutative ideal $\h = \langle e_1, e_2 \rangle$ is a polarization at $\sigma(J)$. 
This follows from the facts that $[\h,\h] = 0$ and its dimension satisfies $\dim \h = \frac{1}{2} \left ( \dim \g - \ind \g \right ) = 2$.
Applying the technique for constructing $\lambda$-representations~\cite{Shi00},  we obtain the following skew-Hermitian operators, which form two operator-irreducible representations parameterized by $J = \pm 1$:
\begin{equation*}
\ell_1 = i J q_2^2,\quad
\ell_2 = i J q_1 q_2,\quad
\ell_3 = - q_2 \frac{\p}{\p q_1} + i J q_1 q_2 \ln q_2,\quad
\ell_4 = - q_2 \frac{\p}{\p q_2}.
\end{equation*}
These operators act on the space of functions of $q = (q_1, q_2)$, where $q_1 \in \mathbb{R}$ and $q_2 \in \mathbb{R}^+$.
Their skew-Hermitian property holds with respect to the inner product~\eqref{InnProd}, defined by the measure $d\mu(q)$ and the corresponding delta-function $\delta(q,q')$ on $Q = \mathbb{R} \times \mathbb{R}^+$:
\begin{equation}
\label{Ex2mu}
d\mu(q) = \frac{dq_1 \wedge dq_2}{q_2},\quad
\delta(q,q') = \delta(q_1 - q'_1) \delta(\ln q_2 - \ln q_2').
\end{equation}

The solution to the system of equations~\eqref{SysEqD} with the initial condition $\D^J_{qq'}(e) = \delta(q,q')$ is given by
\begin{multline}
\label{Ex2D}
\D^J_{qq'}(x) = 
\delta \left ( q_1 - q_1' - q_2 x_3 \right ) 
\delta \left ( \ln q_2 - \ln q_2' - x_4 \right ) \times \\
\times \exp \left [ i J x_1 q_2^2 + \frac{i J q_2}{2} \left ( q_1 + q_1' \right ) \left ( x_2 + x_3 \ln q_2 \right ) \right ].
\end{multline}
Due to the non-unimodularity of the group $G$, the orthogonality and completeness relations~\eqref{OrtConds} and \eqref{ComConds} require modification~\cite{Bre10}.
To formulate them, we introduce auxiliary functions $\tilde{\D}^J_{qq'}(x)$ differing from $\D^J_{qq'}(x)$ by the multiplier $\Lambda(q') = (q'_2)^4$:
\begin{equation*}
\tilde{\D}^J_{qq'}(x) = \Lambda(q') \D^J_{qq'}(x).
\end{equation*}
The modified relations then take the form:
\begin{equation}
\label{Ex2OrthCond}
\int_G \overline{\D^J_{qq'}(x)} \, \tilde{\D}^{\tilde{J}}_{\tilde{q}\tilde{q}'}(x) d\mu_L(x) = 
\delta(q,\tilde{q}) \delta(q',\tilde{q}') \cdot 2\pi^2 \delta_{J\tilde{J}},  
\end{equation}  
\begin{equation}
\label{Ex2CompCond}
\frac{1}{2\pi^2} \sum_{J = \pm 1} \int_{Q \times Q} \overline{\D^J_{qq'}(x)} \, \tilde{\D}^J_{qq'}(\tilde{x}) d\mu(q) d\mu(q') = 
\delta(x,\tilde{x}),
\end{equation}
Here, $\delta_{J\tilde{J}}$ is the Kronecker delta, and $\delta(x,\tilde{x})$ is the delta function on $G$ with respect to the left Haar measure $d\mu_L(x) = e^{4x_4} dx_1 \wedge dx_2 \wedge dx_3 \wedge dx_4$, explicitly given by
\begin{equation*}
\delta(x,\tilde{x}) = e^{-4x_4} \delta(x_1 - \tilde{x}_1) \delta(x_2 - \tilde{x}_2) \delta(x_3 - \tilde{x}_3) \delta(x_4 - \tilde{x}_4).
\end{equation*}
The measure $d\mu(q)$ and the delta function $\delta(q,q')$ on the manifold $Q = \mathbb{R} \times \mathbb{R}^+$ are defined in Eq.~\eqref{Ex2mu}.

The orthogonality and completeness relations~\eqref{Ex2OrthCond} and \eqref{Ex2CompCond} allow us to define the direct and inverse generalized Fourier transforms on $G$ as:
\begin{equation*}
\hat{\psi}(q,q',J) = \int_G \overline{\D^J_{q'q}(x)}\, \psi(x) d\mu_L(x),\quad
\psi(x) = \frac{1}{2\pi^2} \sum_{J = \pm 1} \int_{Q \times Q} \tilde{\D}^J_{q'q}(x) \, \hat{\psi}(q,q',J) d\mu(q) d\mu(q').
\end{equation*}
The presence of the multiplier $\Lambda(q') = (q'_2)^4$ in $\tilde{D}^J_{qq'}(x)$ has an important consequence: under the action of~$\mathcal{F}$, the left-invariant vector fields $\xi_i$ transform according to the rule
\begin{equation*}
\mathcal{F} \circ \xi_i(x,\p_x) = \tilde{\ell}_i(q,\p_q; J) \circ \mathcal{F},
\end{equation*} 
where the operators $\tilde{\ell}_i(q,\p_q; J)$ are related to the original operators $\ell_i(q,\p_q; J)$ by the similarity transformation $\tilde{\ell}_i(q,\p_q; J) = \Lambda^{-1}(q) \ell_i(q,\p_q; J) \Lambda(q)$.
Explicitly, they are given by:
\begin{equation*}
\tilde{\ell}_1 = \ell_1,\quad
\tilde{\ell}_2 = \ell_2,\quad
\tilde{\ell}_3 = \ell_3,\quad
\tilde{\ell}_4 = \ell_4 - 4.
\end{equation*}

Consequently, applying the generalized Fourier transform to the Laplace--Beltrami equation~\eqref{Ex2LBEq} yields the reduced equation
\begin{multline}
\label{Ex2redLBEq}
\redDelta \hat{\psi}(q,q'; J) \defeq 
\left [ 2 \ell_1 \ell_3 + \frac{2}{\beta}\, \ell_2 \left ( \ell_4 - 4 - \alpha \ell_3 \right ) + \frac{1}{\beta} \left ( \alpha \ell_1 +3 \ell_2 \right ) \right ] \hat{\psi}(q,q'; J) = \\
E \hat{\psi}(q,q'; J).
\end{multline}
Explicitly, this equation is a first-order partial differential equation of the form 
\begin{equation}
\label{Ex2redLBEq2}
\left ( Z + V \right ) \hat{\psi}(q,q'; J) = 0,
\end{equation}
where $Z$ is the vector field
\begin{equation*}
Z = \left ( \alpha q_1 - \beta q_2 \right ) \frac{\p}{\p q_1} - q_1 \frac{\p}{\p q_2},
\end{equation*}
and $V$ is the scalar function
\begin{equation*}
V = \frac{\alpha}{2} - i J q_1 \ln q_2 \left ( \alpha q_1 - \beta q_2 \right ) + \frac{i J \beta E}{2 q_2^2} - \frac{5 q_1}{2 q_2}.
\end{equation*}

The vector field $Z$ can be rectified locally by an appropriate coordinate transformation.
As such coordinates, we can take the local coordinates $v$ and $u$ in a neighborhood of the point $q_1 = 0$, $q_2 = - 1/\beta$:
\begin{equation}
\label{Ex2RectCoords}
v = \frac{1}{\lambda_1 - \lambda_2} \ln \left ( \frac{q_1 + \lambda_2 q_2}{q_1 + \lambda_1 q_2} \right ),\quad
u = \left ( q_1 + \lambda_1 q_2 \right )^{\lambda_1} \left ( q_1 + \lambda_2 q_2 \right )^{-\lambda_2},
\end{equation}
where $\lambda_1$ and $\lambda_2$ are the roots of the quadratic equation $\lambda^2 - \alpha \lambda - \beta = 0$:
\begin{equation*}
\lambda_{1,2} = \frac{\alpha \pm \sqrt{\alpha^2 + 4 \beta}}{2}.
\end{equation*}
Here, we assume that $\lambda_1 - \lambda_2 = \sqrt{\alpha^2 + 4 \beta} \neq 0$.
Indeed, it is straightforward to verify that in these coordinates, the vector field simplifies to $Z = \p/\p v$.
Thus, the general solution of the reduced PDE~\eqref{Ex2redLBEq2} have the form
\begin{equation*}
\hat{\psi}(q,q'; J) = C(u, q'; J) e^{-\int V(q(v,u); J) dv},
\end{equation*}
where $q = q(v,u)$ denotes the inverse of the coordinate transformation~\eqref{Ex2RectCoords}, and $C(u, q'; J)$ is an arbitrary functions of the variable $u = u(q_1, q_2)$, the parameters $q' = (q'_1, q'_2)$, and the index $J = \pm 1$.
Although the explicit evaluation of the integral in this solution is cumbersome, it can be expressed in terms of hypergeometric functions.  

Since the function $u = \left ( q_1 + \lambda_1 q_2 \right )^{\lambda_1} \left ( q_1 + \lambda_2 q_2 \right )^{-\lambda_2}$ is an invariant of the vector field $Z$ (i.e., $Z u = 0$), it acts as a symmetry operator for the reduced equation~\eqref{Ex2redLBEq}:
\begin{equation*}
[\redDelta, u] = 0.
\end{equation*}
Clearly, after applying the inverse generalized Fourier transform, the function $u$ becomes an operator acting on functions on the group $G$, which is a symmetry operator for the original Laplace–Beltrami equation~\eqref{Ex2LBEq}.
Up to a multiplicative constant, this symmetry operator admits a formal representation via left-invariant vector fields:
\begin{equation*}
\hat{U} = \xi_1^{\frac{\lambda_2 - \lambda_1}{2}} \left ( \xi_2 + \lambda_1 \xi_1 \right )^{\lambda_1} \left ( \xi_2 + \lambda_2 \xi_1 \right )^{-\lambda_2}.
\end{equation*}
The symbol $\xi_1^\alpha$ denotes the operator obtained by lifting the multiplication operator $\ell_1(q; J)^\alpha$ from $L^2(Q)$ back to $L^2(G)$.
Such an operator is nonlocal and, in general, cannot be expressed as a finite-order differential operator.

Thus, the second example explicitly demonstrates the emergence of nonlocal symmetry operators predicted in section~\ref{sec:3}, confirming that the metric $\G^{(1)}$ indeed belongs to the  class we considered.


\section*{Conclusion}
\addcontentsline{toc}{section}{Conclusion}

In this paper, we have identified and studied a special class of left-invariant pseudo-Riemannian metrics on Lie groups for which the Laplace--Beltrami equation admits exact solutions. 
The main results can be summarized as follows.

First, we established a simple algebraic condition characterizing this class: the existence of a commutative ideal $\h$ in the Lie algebra $\g$ whose orthogonal complement $\h^\perp$ with respect to the metric $\G$ is contained in $\h$. 
This condition, $\h^\perp \subseteq \h$, has several immediate geometric consequences. 
It forces the ideal $\h$ to be at least half the dimension of $\g$ and, in the Riemannian case, forces the Lie algebra itself to be commutative. 
In the Lorentzian case, it permits nontrivial examples where $\h$ is a codimension-one commutative ideal, while for higher indefinite signatures it yields a rich family of non-Riemannian metrics.

Second, we showed that under this condition the Laplace--Beltrami operator acquires a special algebraic structure: in a basis adapted to the ideal $\h$, the components $\G^{ab}$ of the inverse metric vanish whenever both indices lie outside $\h$. Consequently, the operator $\Delta$ becomes linear in the left-invariant vector fields $\xi_{s+1},\dots,\xi_n$ that span a complement to $\h$.

Third, and most importantly, we proved that for this class of metrics the Laplace--Beltrami equation reduces under the generalized Fourier transform to a first-order linear partial differential equation on the homogeneous space $Q = G/P$. 
This reduction is the key to explicit integrability and stands in contrast to the general case, where the reduced equation remains of second order.

Fourth, we demonstrated that this first-order reduced equation can be integrated explicitly by rectifying the characteristic vector field appearing in the reduced operator. 
The general solution is expressed in terms of an arbitrary function of the invariants of this vector field, together with an exponential factor containing an integral of the remaining coefficients.

Fifth, we showed that the invariants of the characteristic vector field give rise, via the inverse generalized Fourier transform, to nonlocal symmetry operators for the original Laplace--Beltrami equation on $G$. 
These operators are generically integro-differential rather than finite-order differential operators, reflecting the non-polynomial nature of the underlying integrals of motion. 
This phenomenon, predicted in the Introduction, distinguishes our class of metrics from those previously studied in the literature, where additional symmetries are typically polynomial in the momenta and correspond to differential operators of finite order.

The theoretical results were illustrated by two concrete examples. 
The three-dimensional Heisenberg group $\H{3}{\R}$, endowed with the Lorentzian metric having a null center, served as a test case: the noncommutative integration method reproduced the general solution obtained by classical separation of variables, thereby validating the consistency of the framework. 
The four-dimensional non-unimodular group of type $\g_{4,7}$, equipped with a metric of signature $(2,2)$ from the canonical family $\G^{(1)}$, provided a nontrivial illustration. 
In this example, the absence of a three-dimensional commutative subalgebra makes classical separation of variables extremely difficult, yet the noncommutative method yielded explicit solutions and revealed the predicted nonlocal symmetry operator $\hat{U}$.

Several directions for future research remain open. 
The most immediate is the algebraic classification of pseudo-Riemannian Lie algebras $(\g,\G)$ satisfying the condition $\h^\perp \subseteq \h$ for some commutative ideal $\h$. 
Such a classification would provide a complete list of Lie groups and metrics to which our method applies. 
Preliminary results indicate that this condition imposes strong restrictions on the Lie algebra structure, and a systematic classification, at least in low dimensions, appears feasible.

A second important direction concerns the functional-analytic aspects deliberately omitted from the present work. 
The solutions we constructed are understood in the distributional sense, and the question of their membership in specific function spaces (such as $L^2(G)$, Schwartz spaces, or Sobolev spaces) depends on the choice of the arbitrary functions appearing in the construction. 
A detailed investigation of these functional properties, possibly leading to spectral decompositions of the Laplace--Beltrami operator for the metrics in our class, would be of considerable interest.

Finally, the appearance of nonlocal symmetry operators raises the question of their algebraic structure. 
Do these operators form a closed algebra, and if so, what is its relation to the original Lie algebra $\g$? 
Understanding this algebra might shed light on the hidden symmetries of the system and could potentially lead to a classification of all symmetry operators, not just those arising from the invariants of the characteristic vector field.

We hope to address some of these questions in future work.

\bibliographystyle{ieeetr}
\bibliography{biblio}

\end{document}